# Scalable single-mode Berkeley Surface Emitting Lasers

Rushin Contractor[1]*, Wanwoo Noh[1]*, Walid Redjem*[1], and Boubacar Kanté[1,2]

[1]Department of Electrical Engineering and Computer Sciences, University of California, Berkeley, California 94720, USA
[2]Materials Sciences Division, Lawrence Berkeley National Laboratory, 1 Cyclotron Road, Berkeley, California 94720, USA
bkante@berkeley.edu
*These authors contributed equally to this work

**The scaling of electromagnetic apertures is a long-standing question that has been investigated for at least six decades but has still not been resolved [1-5]. The size of single aperture cavities, bounded by the existence of higher-order transverse modes, fundamentally limits the power emitted by single-mode lasers or the brightness of quantum light sources. The challenge prompted the investigation of vertical cavity surface emitting laser arrays, multilayered engineered photonic crystals, and more recently supersymmetric structures to promote single-mode operation [6-10]. The fundamental challenge stems from the fact that electromagnetic apertures support cavity modes that rapidly become arbitrarily close with the size of the aperture. The free-spectral range of existing electromagnetic apertures goes to zero when the size of the aperture increases, and the demonstration of scale-invariant apertures or lasers has remained elusive. Here, we report open-Dirac electromagnetic apertures that exploit a subtle cavity-mode-dependent scaling of losses in certain class of apertures with linear dispersion of photons at the center of the Brillouin zone. For cavities with a quadratic dispersion, strongly detuned from the Dirac singularity, the complex frequencies of modes converge towards each other with the size of cavities, making cavities invariably multimode. Surprisingly, for a class of cavities with linear dispersion, we discover that, while the convergence of the real parts of cavity modes towards each other is delayed (it still quickly goes to zero), the normalized complex free-spectral range converge towards a constant governed by the loss rate of distinct Bloch bands. We show that this unconventional scaling of the complex frequency of cavity modes in open-Dirac electromagnetic apertures guarantees single-mode operation of large cavities. We experimentally demonstrate that single-mode lasing of such cavities is maintained when the cavity is scale up in size. We name such sources Berkeley Surface Emitting Lasers (BerkSELs). We further show that the far-field emitted by the proposed single-mode BerkSELs corresponds to a topological singularity of charge two, in full agreement with our theory. Open-Dirac apertures open new avenues for light-matter interaction and basic science in open-wave systems with applications in classical and quantum communications, sensing, and imaging.**

Dirac points are topological singularities that have attracted enormous interests since the discovery of massless Dirac fermions governing electronic transport in graphene [11]. Their occurrence usually indicates an imminent topological transition, and they are at the heart of the physics of topological insulators [12]. In general, wave systems that possess a band structure, such as photonic crystals [13-14], also exhibit Dirac cones (DCs) and it was demonstrated that DCs are universal features that can be systematically implemented by controlling the symmetry of the structure [15-16]. In photonics, DCs have mostly been utilized to demonstrate effective zero-index materials and their properties or to control the dispersion of polaritons [17-22]. Recently, a DC implemented using a three-dimensional structure was theoretically proposed to enable large area single-mode lasing [23-24]. However, the proposed mechanism relied on changes in the real mode spacing that increases slightly by an amount much smaller than the typical gain bandwidth of semiconductors and thus insufficient to enable single-mode operation in lasers.

Surface emitting lasers have played a fundamental role in science and technology since the invention of lasers [1], the demonstration of distributed feedback lasers [2], as well as the invention of vertical cavity surface emitting lasers [3]. Scaling the cavity aperture has since been a challenge. We report, a room temperature scalable open-Dirac lasing aperture that exploits the linear dispersion of photons near the band edge of a photonic crystal, and that can be scaled to large sizes. Our proposed structure is presented in Fig. 1a. It is a photonic crystal (PhC) with a hexagonal lattice. The unit-cell of the PhC is presented in the inset of Fig. 1b. The linear dispersion of Fig. 1c is obtained with a laterally infinite structure (X-Y direction) by overlapping a doubly degenerate modes of the $E_2$ irreducible representation and a mode from one of the B irreducible representations of the $C_{6v}$ point group [15-16] (see Supplementary Information). The degeneracy of the modes at the center of the Brillouin zone is obtained for a critical radius of holes $r_{Dirac}$. An open-Dirac cavity, shown in Fig. 1, is formed by truncating the infinite PhC as a hexagon around a central air-hole with edges normal to the M-directions of the lattice. The entire cavity is suspended in air, and it is connected to the main membrane by six bridges at the corners of the hexagon for mechanical stability (Fig. 1a-b). The structure is fabricated using electron-beam lithography and the PhC is defined by inductively coupled plasma etching, followed by a wet etching step to create the suspended membrane (see Supplementary Information).

To understand the system, we computed modes of open-Dirac cavities of different sizes and for holes radii smaller than, equal to, and greater than the critical radius $r_{Dirac}$. Note that we only present the first three modes of the cavities as modes of higher order have a lower quality factor (Q). The computation was performed using a three-dimensional finite-element solver for the transverse electric polarization that corresponds to the polarization providing the highest gain for the multiple quantum wells used in our work. Figure 2a-c (markers) present the computed frequency shifts of the first three cavity modes. The frequency shifts are computed by comparing cavity modes to the frequency of the $B_1$-mode at the Γ-point for an infinite membrane with holes of the same radius. Figure 2d-f (markers) show the scaling of the Q-factor of the same three modes with increasing cavity sizes. When the radius of holes is not close to $r_{Dirac}$, Cavity Mode 1 asymptotes to the frequency of the fundamental mode at a rate of $k^2$, where k is the wavevector in the cavity. This is shown in Fig. 2g along with the scaling for $r_{Dirac}$ in which case the separation increases to scale at the rate of $k$. Note that Cavity Mode 1 flips from being at a lower frequency than the fundamental mode for $r < r_{Dirac}$ to a greater frequency than the fundamental mode for $r > r_{Dirac}$. We also observe that even for the cavity with linear dispersion, the frequency separation rapidly drops to about a THz when the diameter of the aperture reaches $D=31a$, where a is the size of the unit-cell. The gain spectrum of semiconductors and notably the quantum wells on which the devices were fabricated, spans almost 100 THz, i.e., much larger than real mode spacing. The selectivity of the lasing mode can thus not be enabled by the scaling of the frequency shift afforded by linear dispersion alone as initially claimed [24,25].

We now investigate the quality (Q) factor of our proposed Dirac cavities, with, as example, a hexagonal truncation. The hexagonal truncation of the cavity can serve as selector of the fundamental mode shown by circles on solid lines in Fig. 2. The $B_1$-mode has a six-fold rotational symmetry while the $E_2$-modes only have a two-fold rotational symmetry. Thus, the fundamental mode which is spread out over the entire hexagonal cavity cannot originate from the $E_2$ bands. Higher-order modes correspond to momentum vectors displaced from the Γ-point and can have a two-fold rotational symmetry. Hence, these modes may originate from either the $B_1$ band or the $E_2$ band. We will refer to them as Cavity Mode 1 and Cavity Mode 2 and they are denoted by square and triangle markers in Fig. 2a-f. Figure 2d-f (markers) show that, as expected, the Q-factor of all the modes increases with the size. We also observe that the Q-factor of the fundamental mode ($Q_0$) decreases as the radius of the air-holes increases. This can be attributed to a decrease in the average refractive index of the membrane which reduces the confinement of light. Analogous to the scaling of frequency, we observe that the Q-factor of Cavity Mode 1 asymptotes to $Q_0$ when $r$ is detuned from $r_{Dirac}$ [see Fig. 2d ($r < r_{Dirac}$) and Fig. 2f ($r > r_{Dirac}$)]. Surprisingly, when cavities are tuned to the Dirac point ($r = r_{Dirac}$) higher-order modes do not asymptote to the fundamental mode anymore as seen in Fig. 2e. They lose energy at a rate always higher than the fundamental mode. Unlike the normalized real-free-spectral range that still decays quickly with the size (Fig. 2g), the normalized imaginary-free-spectral range maintain a non-decaying value despite increasing cavity sizes (Fig. 2h), thus guaranteeing single-mode operation for large apertures.

This unconventional scaling of open-Dirac cavities can be theoretically understood based on the mixing of B and E modes that form the Dirac cone. In our open-Dirac cavities that are truncation of the infinite structure with no upper limit for the size, the eigenvalue equation can be expressed as,

$$\begin{pmatrix} -\sigma_k + j\gamma_B & 0 & 0 \\ 0 & -\sigma_k + j\gamma_B & \beta\delta_k \\ 0 & \beta\delta_k & \varepsilon + \sigma_k + j\gamma_E \end{pmatrix} \begin{pmatrix} B_1(\infty) \\ B_1(\infty) \\ E_2(\infty) \end{pmatrix} = \widetilde{\omega}(N) \begin{pmatrix} |0> \\ |1> \\ |2> \end{pmatrix}$$

where the cavity modes $|0>$, $|1>$, and $|2>$ are expressed in the basis of the $B_1$ and $E_2$ unit-cell modes of the infinite system. The finiteness of the cavity introduces an uncertainty in the momentum indicated by $\sigma_k \sim 1/4\pi N$ and higher-order modes lie at $\delta_k = \pi/N$ for a cavity with N unit-cells along the diagonal. The loss rates of the modes scale with the size of the cavity as $\gamma_i = c_i N^{-1} + d_i N^{-2}$ (see Supplementary Information), where $i =$ B or E, $c_i$ and $d_i$ are loss rates introduced due to effects of the boundaries. For a large detuning ε of the $E_2$ modes from the $B_1$ modes, the frequency of cavity mode $|1>$ quadratically asymptotes to the frequency of cavity mode $|0>$ and the field profile at position vector $r$ is $|1> = e^{j\delta_k r}|0>$. The imaginary part of the frequency of cavity mode $|1>$ also rapidly tends to the loss rate of cavity mode $|0>$ and the normalized complex (real and imaginary) free-spectral range thus asymptote to zero. Such cavities are not scale-invariant and become rapidly multimode with the size. However, since the $C_6$ symmetry of the cavity is more favorable for the B modes, we find that $d_E > d_B$, and $c_E > c_B$. Moreover, when ε→0 the fields in cavity mode $|1>$ are a linear combination of both the $B_1$ and $E_2$ unit-cell modes (Fig. 2i). Hence its loss rate is also defined by the loss rate of cavity mode $|2>$ which originates from the $E_2$ mode, and, for $N \to \infty$, the normalized complex free-spectral range tends towards a non-vanishing value of $1 - \frac{c_B}{c_E} \sim 0.8$. Theoretical results, plotted in Fig. 2a-h as continuous lines, are in perfect agreement with numerical simulations (markers). Our open-Dirac cavities are thus scale-invariant and remain single-mode when they are scaled up in size.

To experimentally demonstrate open-Dirac electromagnetic apertures, we characterized fabricated Dirac cavities of aperture diameter $D=19a$ (Fig. 3a), $D=27a$ (Fig. 3e), and $D=35a$ (Fig. 3i), where a is the size of the unit-cell. The cavities were optically pumped at room temperature with a pulsed laser (λ = 1,064 nm, T = 12 ns pulse at a repetition rate f = 215 kHz) and the emission from each aperture was collected through a confocal microscope optimized for near-infrared spectroscopy (see Supplementary Information). The signal was directed toward a monochromator coupled to a InGaAs photodiode for spectral measurements. Figure 3 presents the evolution of the normalized output power as a function of the wavelength and the size of the cavity for unit-cell holes radii smaller than the singular radius $r_{Dirac}$ (b, f, j), equal to $r_{Dirac}$ (c, g, k), and greater than $r_{Dirac}$ (d, h, l). For $D=19a$, cavities are single mode for $r < r_{Dirac}$ (b), $r = r_{Dirac}$ (c), and $r > r_{Dirac}$ (d). For $D=27a$, cavities remain single mode for $r < r_{Dirac}$ (f), $r = r_{Dirac}$ (g), and $r > r_{Dirac}$ (h). This is because these cavities are relatively small. However, when the size of cavities is increased to $D=35a$ or larger, we observe that they become multimode for $r < r_{Dirac}$ (j), remain single mode for $r = r_{Dirac}$ (k), and become multimode for $r > r_{Dirac}$ (l). The Dirac singularity erases higher-order modes in open-Dirac cavities and BerkSELs remain single-mode when the size is increased. It is worth noting that the uniform field profile across the aperture for the fundamental mode (Fig. 2k) depletes gain across the aperture, making it even more difficult for higher-order modes to lase. Single-mode lasing is thus maintained in BerkSELs even for near-damage-threshold pump power (see Supplementary Information). BerkSELs are thus robust to size and pump power density scaling because of the non-vanishing complex free-spectral range and the participation of all resonators in the aperture to the lasing mode. These experiments validate our theory and make BerkSELs the first scale-invariant surface emitting lasers. To further characterize the single-mode lasing of BerkSELs, we present in Fig. 4 the light-light curve (Fig. 4a), the second-order autocorrelation at zero delay $g^2(\tau=0)$ [Fig. 4b] and its linewidth (Fig. 4c). The three different regimes corresponding to spontaneous emission (blue region), amplified spontaneous emission (ASE) [yellow region], and stimulated emission (red region) are observed as the pump power is increased. The second-order autocorrelation function shows a transition from spontaneous emission to amplified spontaneous emission (ASE) as its width drops sharply and the bunching $g^2(0)$ increases. The transition from ASE to stimulated emission is evident from $g^2(0)$ decreasing to unity (Fig. 4b) and the width gradually increasing after the lasing threshold (Fig. 4c), unambiguously proving single-mode lasing action from BerkSELs [26-27]. To confirm that lasing originates from the theoretically predicted B-mode (see Fig. 2), experimental far-fields (Fourier space images) of BerkSELs under optical pumping are presented for cavity sizes of $D$=11a (Fig. 4d), $D$=19a (Fig. 4e), $D$=27a (Fig. 4f), and $D$=35a (Fig. 4g). The six-fold symmetry of the beams match with the far-field obtained from simulations of finite cavities (see Supplementary Information). The far-fields originate from the $B_1$ mode with a topological charge of two (see Supplementary Information). Scaling up the cavity size manifests in a smaller beam divergence as expected. We plotted the measured beam divergence as a function of the size of the cavity. The measured beam divergence matches with our theory, and it scales as $1/D$ where $D$ is the diameter of the cavity in full agreement with theory for modes with a flat envelop fully covering an aperture [28].

We have thus demonstrated the first scale-invariant cavities that remain single-mode when the size of the cavity is increased. These open-Dirac electromagnetic cavities are robust against size scaling. The unconventional scaling stems from a complex free-spectral range that does not vanish with the size of cavities. Around the Dirac singularity, higher-order modes are effectively suppressed by their admixture with more lossy-bands of the structure. The fundamental mode of the cavity has a flat envelop that makes all resonators in the aperture participate in the mode. The fundamental mode thus effectively locks all unit-cells (or resonators) in the aperture in phase, a long-standing challenge. Lasers based on such cavities are surface emitting apertures that we named Berkeley Surface Emitting Lasers (BerkSELs). We have confirmed single-mode lasing from BerkSELs by measuring their second order intensity correlation. We have further confirmed the lasing mode by measuring far-field emissions that agree with our theory. These results demonstrate the fundamental importance of openness for enabling scaling in optics, and, they will have implications in linear and non-linear classical and quantum wave-based systems including atoms, electronics, acoustics, phononics, or photonics based on real or synthetic dimensions. The simplicity of BerkSELs makes them universal apertures, readily relevant to applications including virtual reality systems, lidars, interconnects, manufacturing, defense, or lasers for imaging and medicine.

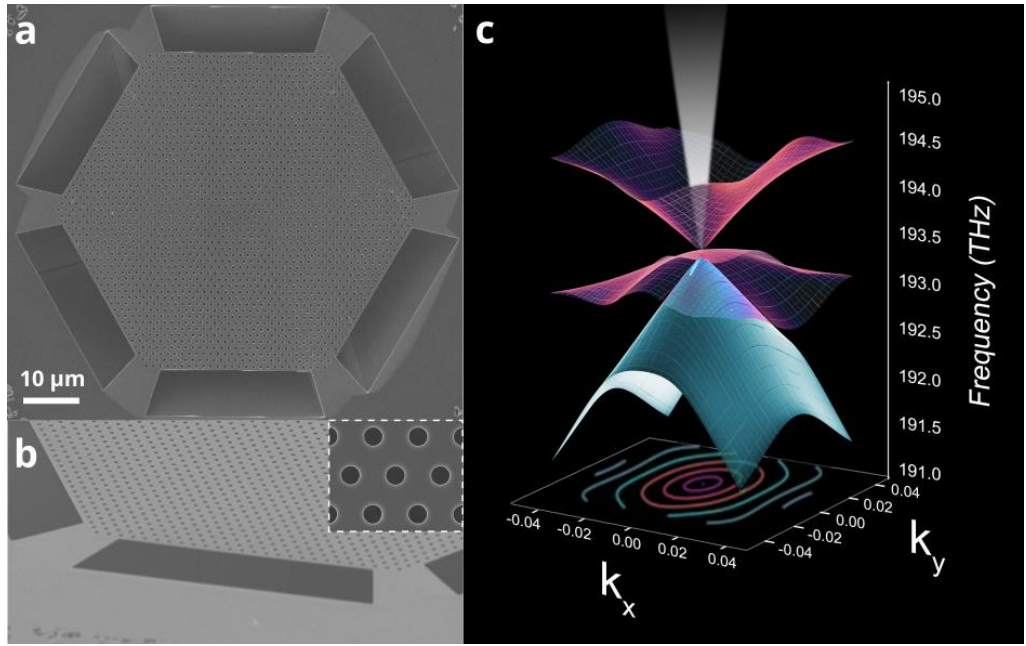

**Figure 1| Scalable open-Dirac electromagnetic cavities. a,** Top view scanning electron micrograph of a hexagonal lattice photonic crystal (PhC) that is truncated to form an open-Dirac electromagnetic cavity. The free-standing structure is suspended via six bridges connecting the main membrane to the substrate along the ΓK direction (Brillouin zone). The cavities are fabricated using electron beam lithography, inductively coupled plasma etching, and wet etching (see Supplementary Information). **b,** Titled view of the cavity showing two bridges, the array of holes, and the PhC-air boundary. The thickness of the membrane is 200 nm, the period of the crystal is 1265 nm, and the radius of holes is used to tune cavities around the Dirac singularity. The inset shows the quality of the nanofabrication with near-perfect circular air-holes interfaces. **c,** Dispersion of the structure displaying a conical degeneracy at 193.5 THz for holes radii of $r_{Dirac}$ = 273 nm. The blue sheet corresponds to the frequency of the $B_1$-mode and the red sheets correspond to $E_2$ modes. The truncation of the crystal, that opens the Dirac cone, is notably chosen to be more favorable for the $B_1$-mode compared to the $E_2$-modes. The iso-frequency contours, projected on the ($k_x$, $k_y$) plane, are sketched together with a representation of laser emission originating from the Dirac point.

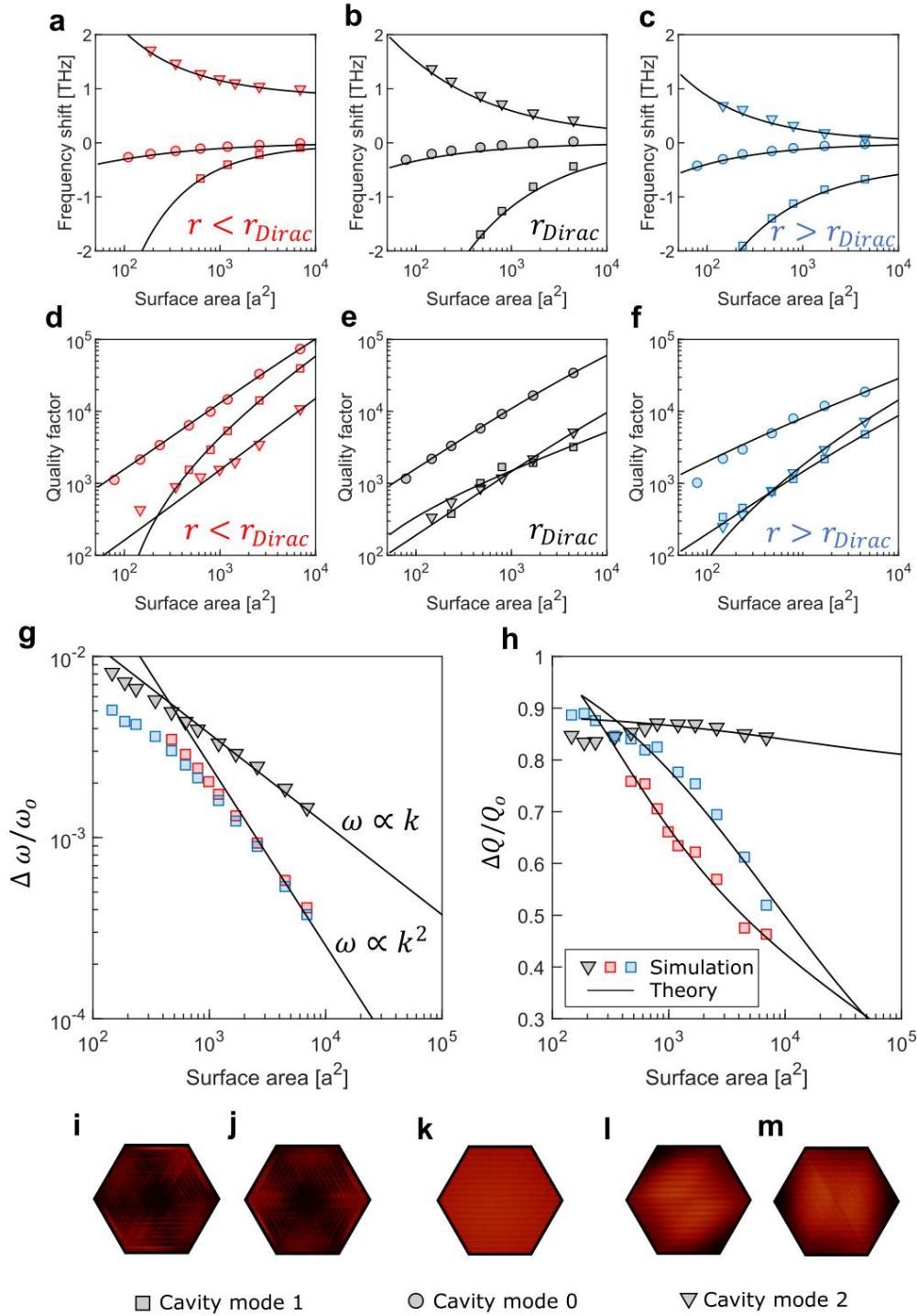

**Figure 2| Complex frequency scaling of open-Dirac electromagnetic cavities.** Frequency shifts of the first three cavity modes for (**a**) $r < r_{Dirac}$, (**b**) $r = r_{Dirac}$, (**c**) $r > r_{Dirac}$, computed by comparing cavity modes to the frequency of the $B_1$-mode at the Γ-point for an infinite membrane with holes of the same radius. Quality factor of the first three cavity modes for (**d**) $r < r_{Dirac}$, (**e**) $r = r_{Dirac}$, (**f**) $r > r_{Dirac}$. **g**, Scaling of the frequency for various radii. When $r$ is detuned from $r_{Dirac}$, the dispersion

is quadratic, and the frequency shift scales as $k^2$. When $r$ is tuned to $r_{Dirac}$ the frequency shift scales as $k$. **h,** Scaling of the quality factor when the radius is detuned from $r_{Dirac}$ and when it is tuned to the singularity. For quadratic dispersion, cavities have an imaginary free-spectral range that vanishes with the size of cavities. Very interestingly, the normalized imaginary free-spectral range does not vanish with the size for our Dirac cavities that can thus be scaled up in size. **(i-j)** Degenerate higher-order cavity mode 1, **k,** fundamental cavity mode, **(l-m)** degenerate higher-order cavity mode 2. Interestingly, the fundamental mode has a flat envelope that synchronizes all emitters in the aperture, corresponding to an effective zero-index mode. For all plots (a-h), markers are numerical simulations and continuous lines are theory based on our model (see Supplementary Information).

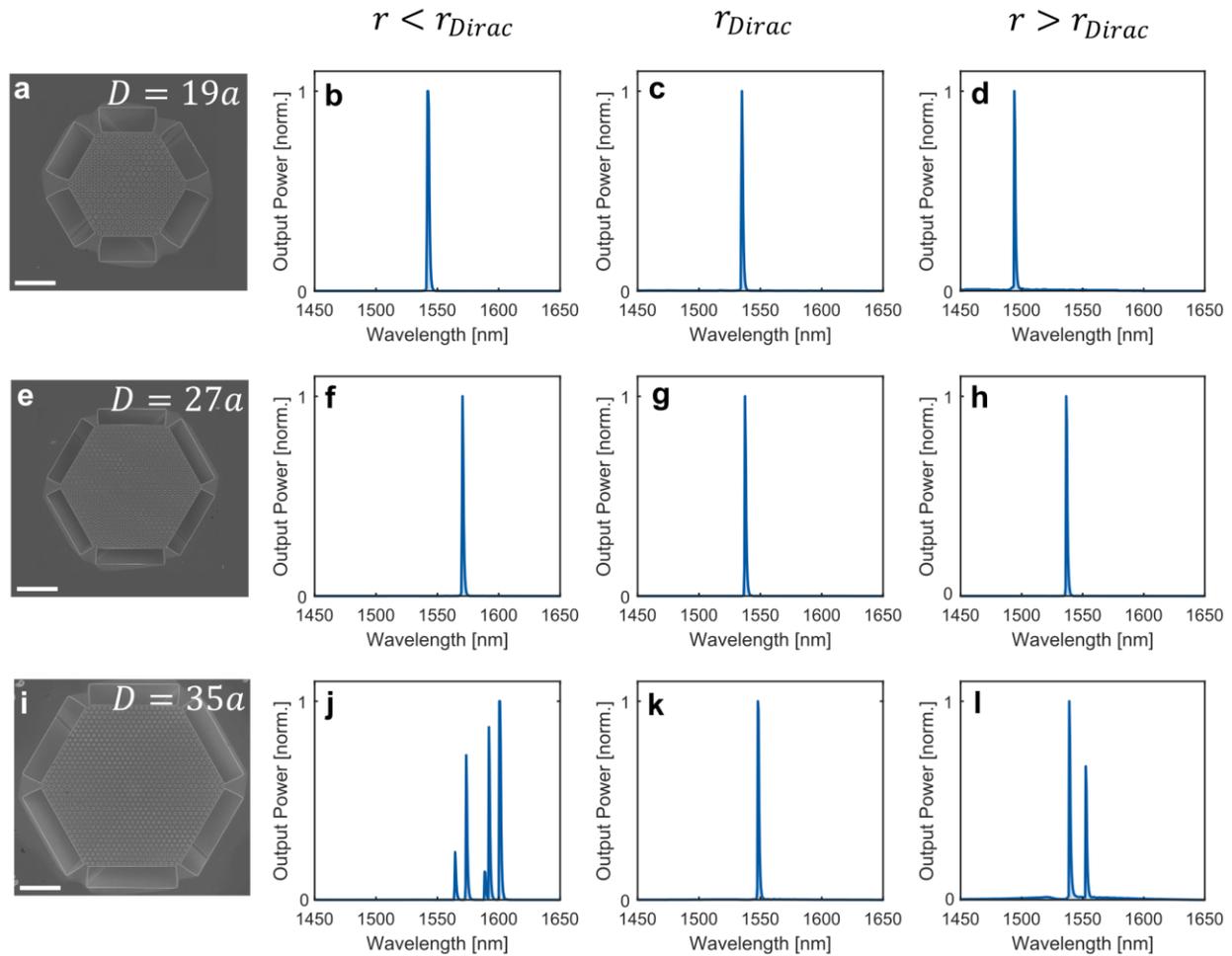

**Figure 3| Lasing characteristics of BerkSELs around the Dirac singularity.** Top-view SEM of fabricated open-Dirac cavities of size $D=19a$ (**a**), $D=27a$ (**e**), and $D=35a$ (**i**), where $D$ is the diameter of the aperture and a is the size of the unit-cell of the photonic crystal. Evolution of the normalized output power as a function of the wavelength and the size of the cavity for unit-cell holes radii smaller than the singular radius $r_{Dirac}$ (**b, f, j**), equal to $r_{Dirac}$ (**c, g, k**), and greater than $r_{Dirac}$ (**d, h, l**). The pump power density is 1.1µW/µm² in all cases. For $D=19a$, cavities are single mode for $r < r_{Dirac}$ (**b**), $r = r_{Dirac}$ (**c**), and $r > r_{Dirac}$ (**d**). For $D=27a$, cavities are single mode for $r < r_{Dirac}$ (**f**), $r = r_{Dirac}$ (**g**), and $r > r_{Dirac}$ (**h**). When the size is increased to $D=35a$, we observe that cavities become multimode mode for $r < r_{Dirac}$ (**j**), remain single mode for $r = r_{Dirac}$ (**k**), and become multimode again for $r > r_{Dirac}$ (**l**). The Dirac singularity erases higher-order modes in open-Dirac cavities and BerkSELs remain single-mode when the size is increased. These experiments validate our theory and make BerkSELs the first scale-invariant surface emitting lasers.

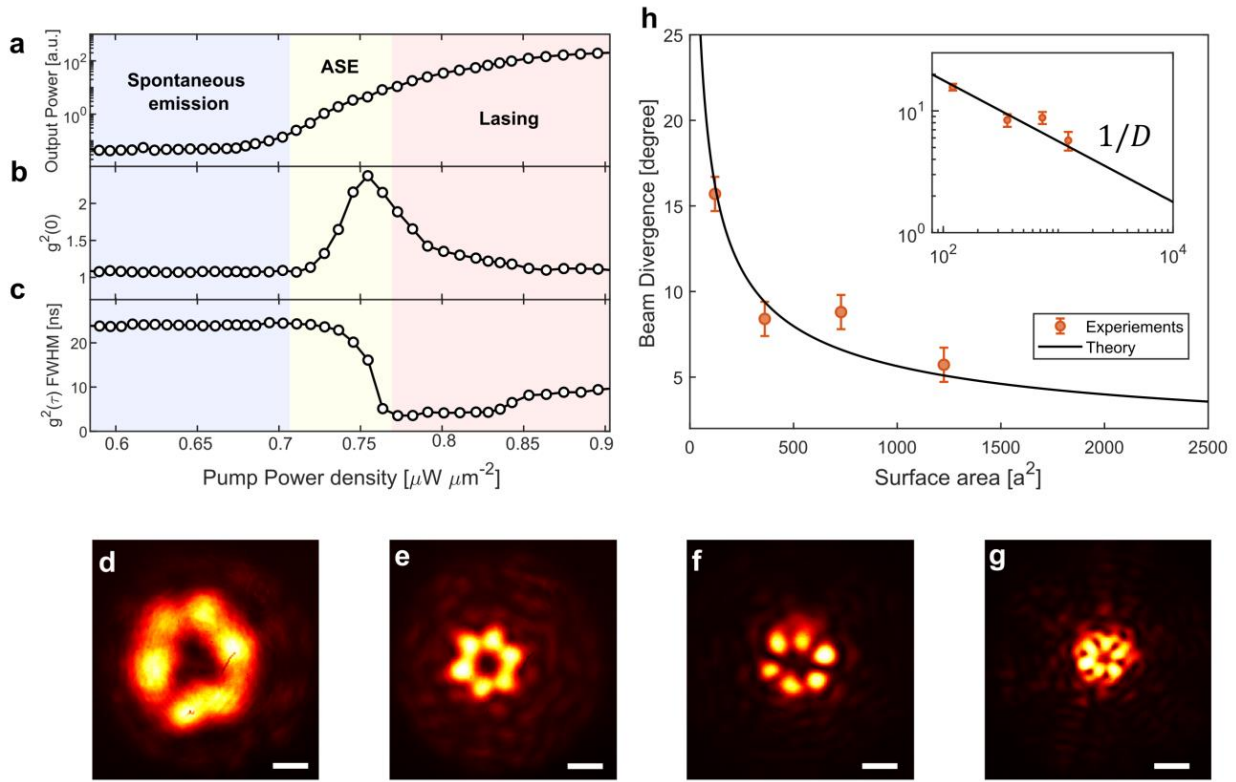

**Figure 4| Photon statistics and far-field of BerkSELs. a,** Emitted output power of a BerkSEL of aperture diameter *D*=35a (where a is the size of the unit-cell) as a function of the pump power density (light-light curve). **b, c,** Second order intensity autocorrelation measurements at zero delay $g^2(0)$ (**b**) and its pulse width (**c**). The pulse width of the second-order autocorrelation function shows a distinct transition from spontaneous emission to amplified spontaneous emission (ASE) as the width drops sharply and then from ASE to stimulated emission as the width gradually increases. These transitions unambiguously demonstrate single-mode lasing from BerkSELs. Experimental far-fields (Fourier space images) of BerkSELs under optical pumping are presented for cavity sizes of *D*=11a (**d**), *D*=19a (**e**), *D*=27a (**f**), and *D*=35a (**g**). The scale bars indicate 10°. Measured and theoretical beam divergence angle as a function of the cavity size. The continuous line is the theoretical prediction and points are experimental data. A good agreement is observed between theory and experiments. The inset shows the same data plotted in log-log scale, demonstrating the 1/*D* scaling of the beam divergence where *D* is the diameter of the aperture (see Supplementary Information). This scaling corresponds to the theoretical limit obtained for modes with a flat envelop fully covering an aperture (see Fig. 2k). Error bars indicate the standard deviation of the beam divergence.